\documentclass[aps,prb,footinbib,superscriptaddress,twocolumn,showpacs,floatfix]{revtex4-1}

\usepackage{amsbsy}
\usepackage{amsmath}
\usepackage{amssymb}
\usepackage{amsfonts}
\usepackage{graphicx}
\usepackage{epsfig}
\usepackage[english]{babel}
\usepackage{color}
\usepackage[normalem]{ulem}
\usepackage{bm}
\usepackage{yfonts}

\begin{document}

\date{\today}

\title{Partial Equilibration of the Anti-Pfaffian edge due to Majorana Disorder}

\author{Steven H. Simon}
\affiliation{Rudolf Peierls Centre for Theoretical Physics, 1 Keble Road, Oxford, OX1 3NP, UK}

\author{Bernd Rosenow}
\affiliation{Institut f\"ur Theoretische Physik, Universit\"at Leipzig, D-04103 Leipzig, Germany}
\begin{abstract}
We consider electrical and thermal equilibration of the edge modes  
 of the Anti-Pfaffian quantum Hall state at $\nu=5/2$ due to tunneling of the Majorana edge mode  to trapped Majorana zero modes in the bulk. Such tunneling breaks translational invariance and allows scattering between Majorana and other edge modes in such a way that there is a 
 parametric difference between the length scales for equilibration of charge  and  heat transport between integer and Bose mode on the one hand, and for thermal equilibration of the Majorana edge mode on the other hand. We discuss a parameter regime in which this mechanism 
 could explain the recent observation of quantized heat transport  [Banerjee et all, Nature  559, 7713 (2018)].  
\end{abstract}


\maketitle

Driven in part by the dream of building a quantum computer\cite{nayakreview} the goal of observing Majorana fermions in condensed matter has been extremely prominent in the last few years\cite{NJPcollection,BeenakkerReview,Franz}, with much effort  devoted  to finding signatures of Majorana zero modes in charge transport \cite{}. In addition, Majorana edge modes existing at the boundary of  a topological state of matter also have a 
 unique signature in heat transport: they contribute one half of the thermal conductance quantum  $K_0 = \kappa_0 T = {\pi^2 k_B^2 \over 3 h} T$ to the thermal Hall conductance, qualitively different from  integer and abelian fractional quantum Hall states, whose thermal conductance  is quantized   in integer  multiples of $K_0$. 
Recently, a half-integer thermal Hall conductance was indeed observed in  the $\nu=5/2$ quantum Hall state \cite{Banerjee2018}, providing evidence for the presence of a Majorana edge mode. 

The thermal Hall conductance is a universal characteristic of a quantum Hall state, since it is independent of details of the edge structure like disorder and interactions. For this reason, it came as a surprise that the experimental value of approximately 
${5 \over 2} K_0$ differs from the theoretical value ${3\over 2} K_0$ for the anti-Pfaffian quantum Hall states, which is expected  to be realized on the $\nu = 5/2$ plateau according to  exact diagonalization in the absence of disorder\cite{RezayiSimon,RezayiRecent}.  Several other possible candidate states do not agree with the experimentally observed thermal Hall conductance either.  While there does exist one proposed state,  the particle-hole symmetric Pfaffian state\cite{Son}, which does have ${5 \over 2} K_0$ thermal Hall conductance, this is  unlikely to be realized in the experiment of Ref.~\onlinecite{Banerjee2018} (See the discussion in Ref.~\onlinecite{SimonPaper}).

The ideal topologically protected thermal Hall conductance can only be observed experimentally when all edge channels are in thermal equilibrium with each other, such that their contributions add up to the universal value, assuming no heat dissipates into the bulk \cite{Steinberg+19}. If a sample is shorter than the thermal equilibration length, then deviations from the universal value are expected. In particular, if the Majorana mode of the Anti-Pfaffian edge is not equilibrated, a 
thermal Hall conductance of ${5 \over 2} K_0$  in agreement with the  experimental observation is expected  \cite{SimonPaper,FeldmanComment,Feldman2}. However, under the assumption that scattering processes leading to equilibration between edge modes are due to charge disorder, it is unlikley that charge transport perfectly equilibrates so as to give perfectly quantized electrical conductance\cite{KaneFisherPolchinski} while at the same time the Majorana mode falls out of  thermal equilibrium \cite{SimonPaper,FeldmanComment,Feldman2}. 

In this letter, we present a different mechanism for edge equilibration which relies on ``Majorana disorder", i.e.~
a coupling between the edge Majorana mode and localized Majorana zero modes in the bulk. 
 In the current discussion the disorder acts nonperturbatively to allow for a new type of scattering process mediated by tunneling to Majorana zero modes on trapped quasiparticles in the bulk. We further give a detailed calculation of the thermal conductance as a function of temperature in reasonable agreement with experiment.

\begin{figure}
%
%
%
%
%
%
%
\includegraphics[scale=.25]{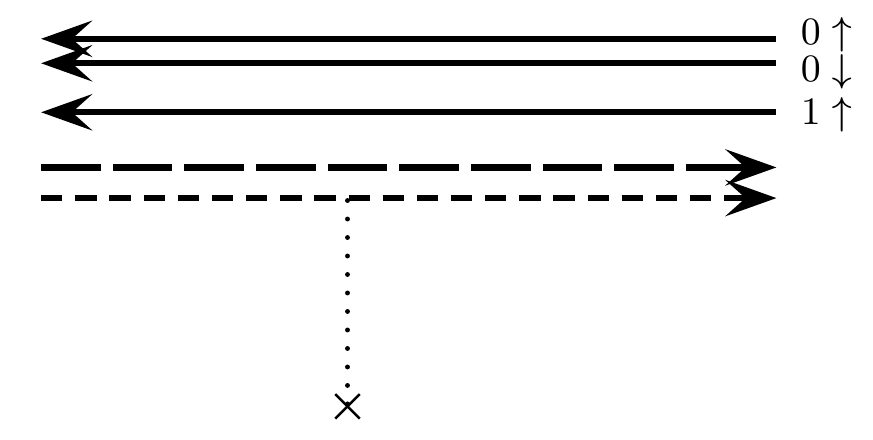}
	\caption{Proposed model of AntiPfaffian edge.    Three integer edge modes (solid) flow downstream $0\!\!\uparrow$, $0\!\!\downarrow$ and $1\!\!\uparrow$.  
	A Bose edge mode (long-dashes) and a Majorana edge mode  (short dashes) flow upstream.  A trapped Majorana zero mode (marked X) is coupled (dots) to the Majorana edge mode. }
	\label{fig:trappedpicture}
\end{figure}

In the absence of disorder, the edge modes of the Anti-Pfaffian comprise three (downstream) integer quantum Hall edge modes, an upstream (reverse-running) bosonic edge mode and an upstream (reverse-running) Majorana edge mode\cite{LeeAntiPfaffian,Levin} (See Fig.~\ref{fig:trappedpicture}).    
As emphasized in Ref.~\onlinecite{FeldmanComment}, one typically expects a momentum mismatch between different edge modes, so that tunneling of an electron between edge modes requires a change in momentum.  Previous discussions have assumed that such a momentum change is provided by charge disorder\cite{SimonPaper,FeldmanComment,Feldman2}.  In the current work we will assume that no such disorder exists, so that this mechanism is inactive.

We thus start by considering a completely smooth edge potential.   Without disorder one might expect neither electrical nor thermal equilibration between edge modes.  However, even for a completely smooth confining potential, since the density is changing gradually, in the bulk there should be trapped quasiparticles or quasiholes near the edge --- each one harboring a Majorana zero mode.  In the absence of disorder these particles will form some sort of Wigner crystal (or glass) minimizing their energies in the smooth potential background as well as minimizing their interaction energies with each other.   Let us assume that some of these quasiparticle locations are not too far from the edge.   We also assume that the Coulomb energy is large enough so that the trapped, charged, particles do not change their positions.   

Generically, there will be coupling of the trapped Majorana zero mode to the Majorana edge mode as shown in Fig.~\ref{fig:trappedpicture}.   Such coupling of the edge to a trapped Majorana has been analyzed in a number of different contexts before\cite{Fendley,Rosenow2,Rosenow1,Bishara,Roising}.    The result of such a coupling is to produce an energy dependent scattering phase shift to the edge Majorana of the following form
\begin{equation} 
\label{eq:scatteringphase}
  e^{i \varphi(E)} = \frac{E + i E_{coupling}}{E - i E_{coupling}}
\end{equation}
where $E_{coupling}$ is the strength of the coupling between the trapped Majorana mode and the edge (See supplement section \ref{sec:phaseshift} for rederivation of this result\cite{Supplement}).

The key here is to realize that at energies high compared to the coupling energy, the Majorana edge mode is undisturbed by its coupling to the trapped mode ($\varphi$ is close to zero).   However, at low energies compared to the coupling energy, the Majorana mode is maximally phase shifted by an angle of $\pi$.  
In particular, for an edge Majorana with wavevector $k$, such that $E=vk \ll E_{coupling}$ the wavefunction takes the form $e^{i k x}$ for $x<x_0$ and $-e^{i k x}$ for $x>0$.   This function has Fourier modes $\sim e^{i (q - k) x_0}/(q - k)$, thus it allows overlap of this Majorana edge mode with other edge modes even with substantial momentum mismatch.  Thus, we should expect there should be scattering into the Majorana edge mode at energies less than $E_{coupling}$ but not at energies much greater than $E_{coupling}$.     

Suppose further that the coupling energy happens to be somewhat smaller than the temperature.   In this case we have a mechanism by which scattering of charge occurs only when the energy of the Majorana is sufficiently low, thus keeping the heat from being transferred to the Majorana mode --- potentially achieving charge equilibration without thermal equilibration. 

Let us now be more precise about the details of the scattering model we solve.    We consider scattering to a single integer mode ($1\!\!\uparrow$ in the  figure) which we write using fermionic fields $\{\psi(x),\psi^\dagger(x') \} = \delta(x-x')$.  The Majorana edge mode is $\xi_0$, and we will use a convenient representation\cite{LeeAntiPfaffian,Levin} of the Bose mode in terms of two Majorana operators $\xi_1$ and $\xi_2$.    These Majorana fields are self conjugate $\xi_\alpha^\dagger = \xi_\alpha$ and have Fermionic anticommutations $\{ \xi_\alpha(x), \xi_\beta(x')\} = \delta_{\alpha \beta} \delta(x - x')$.   The Hamiltonian of the edges is then given by
\begin{equation}
 H_0 = i \! \! \int dx  [v_i \psi^\dagger(x)\partial_x \psi(x) + \! \! \sum_{\alpha=0,1,2}\! \! \frac{v_\alpha}{2}  \xi_\alpha(x) \partial_x \xi_\alpha(x) ]
\end{equation}
where $v_i < 0$ is the integer mode velocity, $v_0>0$ is the Majorana $\xi_0$ velocity, and $v_1 = v_2 = v_b>0$ is the Bose velocity.   In the presence of large disorder scattering Refs.~\onlinecite{Levin,LeeAntiPfaffian} consider a fixed point where $v_0=v_1 = v_2$.   However, here we are assuming low disorder limit and generally we expect that the Majorana velocity $v_0$ is somewhat less\cite{RezayiWanYang}  than the Bose or integer velocities $v_b$ and $v_i$.   On the right hand side we assume a reservoir at temperature $T$ and voltage $0$ whereas on the left we assume reservoir with temperature $T+\Delta T$ and voltage $V$. 

In addition we add an interaction induced scattering term to allow an electron to scatter from the integer to the fractional edges.  
This is of the form
\begin{equation}
 H_1 = \alpha \int dx  \, e^{i p x} \, \psi^\dagger(x) \xi_0(x) \xi_1(x) \xi_2(x) + \mbox{h.c.}
\end{equation}
where $\alpha$  is a coupling constant with dimensions of velocity which should be roughly on the order of the edge mode velocity (to be detailed further below and in the supplement\cite{Supplement}),  and $p$ is the wavevector mismatch between the integer and fractional modes (assumed to be on the order of the inverse magnetic length).     Note that here the electron in the fractional edges is made of a product of the three Majoranas.   In the absence of additional disorder, due to the wavevector mismatch $p$,
there can be no scattering at low voltage and low temperature difference between the edge modes. 

Finally, we add the single Majorana impurity $\gamma_{qp}$ zero mode ($\gamma_{qp}^2=1$ and $\{\gamma_{qp}, \xi_j(x)\} = 0)$, via the Hamiltonian
\begin{equation}
H_2 = i  \lambda \, \gamma_{qp} \,  \xi_0(x_0)
\end{equation}
where $x_0$ is the position of the coupling, and $\lambda$ is the coupling constant.   If we start by ignoring the Bose and integer mode,  it is easy to show that the phase shift 
to the $\xi_0$ mode due to the coupling $H_2$ is given by Eq.~\ref{eq:scatteringphase} where $E_{coupling}  = \lambda^2/v_0$ (See supplement section \ref{sec:phaseshift} for detailed derivation\cite{Supplement}).

We now sketch the calculation of the tunneling current between the integer and fractional edges.  Full details are given in the supplement\cite{Supplement} section \ref{sec:MASC}.   With the help of Fermi's golden rule we write the tunneling current through the impurity as\cite{KaneFisher}
\begin{widetext}
\begin{eqnarray}
	J^\alpha 
&\sim &   \int \! \! dx \! \int \! \! dx'  \! \!\! 	\int\! \!  dE \, X^\alpha \left[ e^{i p (x - x')}   G_<^L(E,x',x)  G_>^R(E+eV,x',x) -   e^{-i p (x - x')}   G_>^L(E,x',x) G_<^R(E+eV,x'x)\right]_,	
\label{tunneling.eq}
\end{eqnarray}
\end{widetext}
where $\alpha=e$ or $E$ (for charge current or energy current), $X^e = -e$ and $X^E = E$ with $p$ the momentum mismatch between the right- and left-moving edges and $V$ the voltage difference.    On the left moving 
integer edge,  
%
$
G^L_{>,<}(E,x^\prime,x) \ \sim  e^{\pm i (E/v_i) (x - x^\prime)} n_F(\mp E), 
$
%
where $n_F(E) = 1/(1 + e^{\beta E})$ denotes the Fermi distribution, and $\beta = 1/ k_B T$. The right moving electron Green's function can be expressed as a convolution  of Bose and Majorana Green's functions
%
\begin{equation} \nonumber
G^R_{>,<}(E,x,x') \sim \!\!\int \!\!dE' \,\, G_{>,<}^b(E-E',x',x)  G_{>,<}^\xi(E',x',x)
\end{equation}
%
Here, 
%
$
G_{>,<}^b(E,x,x') \sim  \mp E n_B(\mp E) e^{\mp i (E/v_b) (x - x') }.
$
%
The Majorana Green's function in the absence of the impurity is
%
$
G_{\substack{>,<}}^{\xi,0}(E,x,x') \sim n_F(\mp E) e^{\mp i (E/v_0) (x - x') }  $.
%
In the presence of an impurity, the Majorana Green's function is given by 
%
$
G_{>,<}^\xi(E,x,x')  = G_{>,<}^{\xi 0}(E,x,x') F(E,x,x')
$
%
with a phase shift from the impurity at position $x_0$
%
\begin{equation}
F(E,x,x')= \left\{ \begin{array}{ll}   e^{i \varphi(E)}   & x > x_0 > x' \\ e^{-i \varphi(E)}  & x < x_0 < x'  \\ 1 & \mbox{otherwise} \end{array}  \right. 
\end{equation}
where $\varphi(E)$ is given by Eq.~(\ref{eq:scatteringphase}). 
%

Evaluating the tunneling current Eq.~(\ref{tunneling.eq}) using the above Green's functions (See supplement\cite{Supplement} section \ref{sec:MASC} for details) we obtain results  in line with the  expectations described earlier.   While the most general analytic expression can be somewhat complicated, we can more easily examine the limit of  very weak coupling $E_{coupling}$, and with the assumption that the wavevector mismatch $p$ between the Bose mode and the integer mode is larger than $T/v_0$.   In this limit the electrical conductance from the integer to fractional (combination of Bose and Majorana) modes is given by  
\begin{equation}
 G = \frac{\pi |\alpha|^2 E_{coupling} T}{8 |v_i| v_b^2 v_0 p^2} \, G_{0.}
 \label{eq:Gresult}
\end{equation}
with $G_0 = e^2/h$.   The thermal conductance in this limit is more complicated since the three edge modes can have three different temperatures.   We find the corresponding thermal conductances to be 
\begin{eqnarray} \label{eq:Jresults}
K^{ib} &=&  (k_B/e)^2 (\pi^2/2) T G \\
K^{im} & =& \epsilon K^{ib} \nonumber \\
K^{bm} &=&  2 \epsilon K^{ib} \nonumber \\
 \epsilon &=&  (32/(9 \pi^3)) E_{coupling}/T \approx 0.1 E_{coupling}/T \nonumber
\end{eqnarray}
    where $i$, $b$ and $m$ indicate the integer, Bose and Majorana edge modes.  (For example, the thermal current between the integer and Bose mode is $K^{ib}$ times the temperature difference between these two modes).  There are no thermo-electric couplings due to the particle-hole symmetry of the model \cite{KaFi96}, and the influence of Joule heating on edge temperature and shot noise \cite{Park+19} is neglected due to the leading order expansion in the tunnel coupling $\alpha$.  
    
Assuming the coupling $E_{coupling}$ is sufficiently smaller than $T$, the parameter $\epsilon$ will be small and the thermal conductance into the Majorana mode will be much less than that into the Bose mode.     Thus one should have a regime where there is electrical equilibration, and the Bose mode is fully thermally equilibrated, but the Majorana mode is not. 

Let us assume that heat is not flowing into the Majorana mode.     If we further assume that the $1\!\!\uparrow$ integer mode does not mix with the other integer edge modes, then, this mode, along with the Bose mode form a system of two counter-propagating edges.    This is similar to the case of Ref.~\onlinecite{KaneFisherPolchinski}.  It is expected in such cases that thermal equilibration is diffusive, and the system may not fully equilibrate.   This physics is certainly seen in experiment\cite{BanerjeeAbelian} at $\nu=2/3$, and, as pointed out in Ref.~\onlinecite{Feldman2}, is likely also occurring in experiment\cite{Banerjee2018} at $\nu=2+2/3$ with a similar assumption that the outer two integer edge modes are not mixing with the other modes.   Since the conductances are dropping proportional to $T$ at low temperature, we should expect that equilibration should be particularly bad at low temperature.   
Should the Bose mode go out of thermal equilibrium with the integer mode, the measured thermal conductance should rise\cite{SimonPaper}, which is precisely what is observed in experiment.

We remind the reader that the conductances and thermal conductances calculated so far are conductances between edge modes through a single scattering center.   The electrical conductance between edge modes per unit length is given by $\tilde G = n_{imp} G$ where $n_{imp}$ is the number of scatterers per unit length.    We can define a characteristic charge equilibration length $
 \ell_{e}^b =G_0/\tilde G. 
$
To determine the total electrical conductance along the edge we use the relationship between current and chemical potential being given by $j_\alpha = G_\alpha \delta \mu_\alpha$ with $G_i=G_0$ and $G_b =G_0/2$.  We then include scattering between the two edges via $\partial_x j_{i,b} = \pm \tilde G (\delta \mu_i - \delta \mu_b)$. The solutions of these equations show us that corrections to the quantized electrical conductance will be order $e^{-L/\ell_e^b}$ with $L$ the length of the edge.  (See details of derivation in supplement\cite{Supplement} section \ref{sec:edgeequil}).   Since the quantization of electrical conductance is fairly good, we must assume that $L/\ell^b_e \gg 1$.  

Similarly to the electrical case the thermal conductances per unit length between edge modes $\alpha,\beta\in \{i,b, m\}$ are given by $\tilde K^{\alpha\beta} = n_{imp} K^{\alpha \beta}$ giving a characteristic thermal  length for equilibrating the Bose and integer modes given by 
$
 \ell_{q}^b =K_0/\tilde K^{ib} = 2 l_e^b/3 
$
with $K_0  = (\pi^2/3) T k_B^2/h$.   The thermal current along an edge is given by $J_\alpha = c_\alpha K_0 \, \delta T_\alpha$ where $\alpha = \{i,b,m\}$ and $c_{\alpha} = (-1,1,1/2)$ is the signed central charge of the different edge modes.    We then include scattering between edges via $\partial_x J^\alpha = - \sum_\beta  \tilde K^{\alpha \beta} \delta T_\beta$ with $\tilde K^{\alpha \alpha}$ defined to give energy conservation 
$\sum_\beta \tilde K^{\alpha \beta} = 0$. 
Here,  because we have counter-propagating modes\cite{KaneFisherPolchinski,BanerjeeAbelian,Nosiglia+18}, as 
in the case of $\nu=2/3$,  corrections to the measured quantized thermal conductance will be algebraic.   The solution of this system of equations (detailed in supplement\cite{Supplement} section \ref{sub:thermaledge})  gives us the net thermal conductance of the edge
(including $2K_0$ from the lowest Landau level edges) 
\begin{equation}
\label{eq:KresC}
 K/K_0  = 2.5 + \frac{2}{1 + A T }   - \epsilon \, C(A T) 
\end{equation}
where $A = L/(l_q^b T)$ is a temperature independent constant and where  $\epsilon = (32/(9 \pi^3)) (E_{coupling}/T)$ is the above discussed small parameter which we can approximate as zero if the Majorana mode is decoupled from the integer and Bose modes. For $x \gg1$, we have $ C(x)\approx x$, multiplied by a small and finite $\epsilon$.  Note that we expect the thermal equilibration length for the Majorana mode to scale as $l_q^B/\epsilon$ which can be much longer than the length of the sample.   

\begin{figure}
\includegraphics[scale=.5]{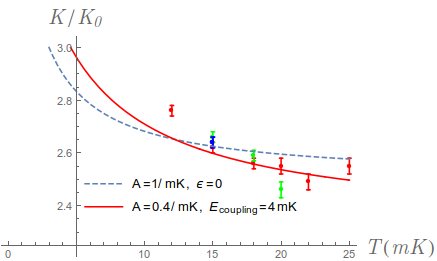}
\caption{Thermal conductance as a function of temperature.  Points are experimental data from Ref.~\onlinecite{Banerjee2018}.   Red, green, blue points are $\nu=2.50,2.49,2.51$ respectively.  The dashed curve is the $E_{coupling}\rightarrow 0$ limit while keeping finite $A=1/$mK.    The solid curve is  Eq.~\ref{eq:KresC} given in the text with $A=0.4/$mK and $E_{coupling} =4$mK.}
\label{fig:data}
\end{figure}

In Fig.~\ref{fig:data} we show example results of this theory compared against experimental data from Ref.~\onlinecite{Banerjee2018}.    The two curves have values of $A$ fit to the data given a fixed value of $E_{coupling} = 0$ or $4$\rm{mK}, showing that the curve shape is relatively independent of $E_{coupling}$.    Note that for all plotted values of $T$ we have $L/l_q^b = (2/3) (L/l_q^c)$ substantially greater than 1.  Thus the measured electrical conductivity will be well quantized. 

One possible concern with our model is that the coupling $H_2$ between the isolated quasiparticle and the edge is assumed to occur at one point $x_0$.  The fact that it is a point coupling is responsible for the appearance of arbitrarily large Fourier modes being active.   More realistically the coupling will be smeared out somewhat.    The tunneling from a Majorana impurity to the edge should be exponential with some decay length $\zeta$. 
If the impurity is a perpendicular distance $R$ from the edge, then the smearing of the coupling along the edge should be roughly $\sim \exp(-\sqrt{R^2 + x^2}/\zeta) \approx e^{-R/\zeta} e^{-x^2/(2 R \zeta)}$ with $x$ the distance along the edge, giving a smearing over a length scale on order $w \approx  \sqrt{R \zeta}$ thus preventing the above described scattering mechanism from being effective if the wavevector mismatch is $p \gtrsim w^{-1}$.       We can use an estimate of $\zeta \approx 1.15 \ell_B$ from prior numerical work\cite{baraban,Supplement}, so that we also have $E_{coupling} \approx 1{\rm K} \, e^{-R/\zeta}$.    Given that we want $E_{coupling}$ in the mK range, we estimate $R \approx 6\ell_B$ thus bounding $p \lesssim 0.3/\ell_B$.  See supplement  section \ref{sub:suppedge} for more details\cite{Supplement}.  

Variants of this model can be imagined where the edges have a larger momentum mismatch $p$, and also charge disorder is present.   Let us assume however, that the disorder wavevector is not as large as  $p$.  In this case scattering can not occur due to this disorder wavevector alone.  However one can consider a situation where scattering can occur if the Majorana impurity mechanism provides some of the momentum and the disorder provides the remainder.   A detailed calculation of this more complicated mechanism is beyond the scope of this work, but we expect that very similar physics will result.

We now turn to the physical parameters which will give us this desired value of $A = (3 \pi/16) |\alpha|^2 E_{coupling} n_{imp} L / (|v_i| v_b^2 v_0 p^2) \approx 0.4/$mK used in the above figure (we need $A$ to be not too much less than 0.4/mK so that the electrical conductivity is well quantized at experimental temperatures).  Let us assume the following reasonable parameters:  velocity $v_i=v_b=10^6$ cm/sec for the integer and Bose modes and $v_0=10^5$ cm/sec for the Majorana edge mode.    The coupling constant $\alpha$ also has dimensions of velocity and should be roughly on the same scale.  In the supplement we detail why a good estimate of this parameter should be given by $\alpha^2 = \pi^2 v_b \sqrt{v_b v_0}$.  We take $p = 0.1/\ell_B$, and in the experiment  $\ell_B = 16$nm and $L=150 \mu$m.  Finally we choose  $E_{coupling}$ to be 4 mK $\ll T$ as given in the figure.  In order to have $A \approx 0.4/$mK
this would require one impurity every 120 nm $\approx 8 \ell_B$.  Note in addition that $A$ scales as the inverse square of both $p$ and the velocities, so that a small reduction in either would allow a much lower density of impurities.     We emphasize that there have been no detailed simulations of the AntiPfaffian edge, and it is possible that the edge potential is strongly screened by the outer edge modes resulting in edge velocities being somewhat smaller than in outer edge modes.



To summarize, we have provided a detailed mechanism that potentially explains the observation of the $K/K_0 \approx 2.5$ from Ref.~\onlinecite{Banerjee2018}, by showing how the Majorana edge mode can remain out of thermal equilibrium, despite the fact that all of the edge modes are in electrical equilibrium.   Further we show how the same mechanism can roughly explain the temperature dependence of the experimental data.

Acknowledgements: SHS would like to thank Bertrand Halperin for helpful conversations.  SHS was supported by EPSRC grant EP/N01930X/1. Statement of compliance with EPSRC policy framework on research data:
This publication is theoretical work that does not require
supporting research data.  BR was supported by DFG grant RO 2247/8-1.

\bibliographystyle{apsrev4-1}
\bibliography{test}

\clearpage

\setcounter{equation}{0}
\setcounter{figure}{0}
\setcounter{table}{0}
\setcounter{page}{1}
\makeatletter
\renewcommand{\theequation}{S\arabic{equation}}
\renewcommand{\thefigure}{S\arabic{figure}}
\renewcommand{\bibnumfmt}[1]{[S#1]}

\onecolumngrid

\begin{center}
{\bf \Large SUPPLEMENTARY MATERIAL}

\end{center}

\section{Majorana Assisted Scattering Calculation}

The purpose of this part of the supplement is to derive Eq.~\ref{eq:Jresults},\ref{eq:Gresult} of the main text.   This is a ``standard" but tedious calculation using the bosonized description of the edge[S1].  While some of the formulae appear messy, the procedure is actually quite straightforward.   We include quite a bit of detail for added clarity.

\label{sec:MASC}

\subsection{Tunneling Formalism}

We begin by deriving a general formula for tunneling between two systems $R$ and $L$ with corresponding Hamiltonians $H_R$ and $H_L$.   The full Hamiltonian is of the form
$$
 H = H_L + H_R + \hat T + \hat T^\dagger
$$
where the tunneling term $\hat T$, the tunneling of a single electron from left to right, can be treated as a perturbation.

We will assume that $\hat T$ is a perturbation so  
both the left and right halfs can be described with density matrices $\rho_L$ and $\rho_R$ and the state of the full system is a simple product $\rho = \rho_L \otimes \rho_R$. 

Similarly the unperturbed eigenstates of the system can be described as simple direct products
$
 |a.b\rangle = | a_L \rangle \oplus |b_R \rangle = |a_L b_R\rangle
$
with corresponding eigenenergies
$
 E_{a,b} = E_{a_L} + E_{b_R}
$
where 
$H_L |a_L \rangle = E_{a_L} |a_L \rangle$ and $H_R |a_R \rangle = E_{a_R} |a_R \rangle_. 
$


Setting $\hbar=1$ throughout, the tunneling rate from Fermi's golden rule is is given by
$$
 \Gamma= {2 \pi} \sum_{i,f} 
|\langle f | \hat T | i \rangle|^2  \delta(E_i - E_f) P(i)
$$
with $|i\rangle$ the inital and $|f\rangle$ the final state (of the entire system) where here $P(i)$ is the probability of the initial state occurring.   If there is a voltage difference between the two sides, we can just add that into the argument of the delta function.

%

The net electrical current flowing from the left to the right can then be written as 
\begin{eqnarray*}
  J^e &=&  {(-2 \pi e)}\sum_{i,f} 
  |\langle f_L f_R | \hat T | i_L i_R \rangle|^2  \delta(E_{iL} + E_{iR} - E_{fL} - E_{fR} + e V) P(|i_L i_R\rangle)   \\
  	&-&  {(-2 \pi e) } \sum_{i,f} 
  	|\langle f_L f_R | \hat T^\dagger | i_L i_R \rangle|^2  \delta(E_{iL} + E_{iR} - E_{fL} - E_{fR} -  e V) P(|i_L i_R\rangle)   	
\end{eqnarray*}
The energy current, on the other hand, is
\begin{eqnarray*}
	J^E &=&  {2 \pi} \sum_{i,f}  (E_{iL} - E_{fL})
	|\langle f_L f_R | \hat T | i_L i_R \rangle|^2  \delta(E_{iL} + E_{iR} - E_{fL} - E_{fR} + e V) P(|i_L i_R\rangle)   \\
	&+& {2 \pi}\sum_{i,f} (E_{iL} - E_{fL})
	|\langle f_L f_R | \hat T^\dagger | i_L i_R \rangle|^2  \delta(E_{iL} + E_{iR} - E_{fL} - E_{fR} - eV) P(|i_L i_R\rangle)   	
\end{eqnarray*}
Writing the delta function as an integration over energy, we get
\begin{eqnarray*}
	J^e &=&  {-e}  \int dt 	\sum_{i,f} P(|i_L i_R\rangle)  e^{it (E_{iL} + E_{iR} - E_{fL} - E_{fR})} 
		\left[ e^{it eV} 	|\langle f_L f_R |\hat T  | i_L i_R \rangle|^2 - e^{-i t e V} 	|\langle f_L f_R | \hat T^\dagger | i_L i_R \rangle|^2
	\right] 
	\\
	J^E &=&    \int dt 	\sum_{i,f} (E_{iL} - E_{fL}) P(|i_L i_R\rangle)  e^{it (E_{iL} + E_{iR} - E_{fL} - E_{fR})}  
		\left[ e^{it eV} 	|\langle f_L f_R |\hat T  | i_L i_R \rangle|^2+ e^{-i t e V} 	|\langle f_L f_R | \hat T^\dagger | i_L i_R \rangle|^2
	\right]
\end{eqnarray*}
This can be rewritten using time dependent operators as 
\begin{eqnarray*}
	J^e &=&  {-e}  \int dt 	\sum_{i,f} P(|i_L i_R\rangle) 
		\left[ e^{it eV} 	\langle i_L i_R | T^\dagger(t) | f_L f_R \rangle \langle f_L f_R |\hat T(0)  | i_L i_R \rangle - e^{-i t e V} 	\langle i_L i_R | T(t) | f_L f_R \rangle \langle f_L f_R |\hat T^\dagger(0)  | i_L i_R \rangle \right] \\ 
	 &=&  -e  \int dt 	\sum_{i} P(|i_L i_R\rangle)   	\left[ e^{it eV} 	\langle i_L i_R | \hat T^\dagger(t)  \hat T(0)  | i_L i_R \rangle - e^{-i t e V} 	\langle i_L i_R | 
	 \hat T(t) \hat T^\dagger(0)  | i_L i_R \rangle \right]  
	\end{eqnarray*}
and
\begin{eqnarray*}
	J^E &=&   \int dt 	\sum_{i,f} (E_{iL} - E_{fL}) P(|i_L i_R\rangle)  
		\left[ e^{it eV} 	\langle i_L i_R | T^\dagger(t) | f_L f_R \rangle \langle f_L f_R |\hat T(0)  | i_L i_R \rangle + e^{-i t e V} 	\langle i_L i_R | \hat T(t) | f_L f_R \rangle \langle f_L f_R |\hat T^\dagger(0)  | i_L i_R \rangle \right] \\
	&=&   \int dt 	\sum_{i,f}  P(|i_L i_R\rangle)  
	 	\left[ e^{it eV} 	\langle i_L i_R | T^\dagger(t) | f_L f_R \rangle \langle f_L f_R |[\hat T(0),H_L]  | i_L i_R \rangle + e^{-i t e V} 	\langle i_L i_R | \hat T(t) | f_L f_R \rangle \langle f_L f_R | [\hat T^\dagger(0),H_L]  | i_L i_R \rangle \right] \\
	&=&   \int dt 	\sum_{i}  P(|i_L i_R\rangle)   	\left[ e^{it eV} 	\langle i_L i_R | T^\dagger(t) [\hat T(0),H_L]  | i_L i_R \rangle + e^{-i t e V} 	\langle i_L i_R | \hat T(t)  [\hat T^\dagger(0),H_L]  | i_L i_R \rangle \right] 	
\end{eqnarray*}

The form of the tunneling is given by
\begin{equation}
\label{eq:Tform}
\hat T = g \int dx \, \psi_{eR}^\dagger(x) \psi_{eL}(x)  e^{i p x}
\end{equation}
where $\psi_{eL}^\dagger$ and $\psi_{eR}^\dagger$ are electron creation operators on the left and right side respectively, $p$ is the wavevector mismatch associated with the tunneling, and $g$ is a coupling constant.   This constant $g$ differs from the coupling constant $\alpha$ used in the main text by a dimensionful cutoff length scale, described below.   We then obtain
\begin{eqnarray*}
J^e &=&  -e |g|^2 \int dt \int dx \int dx' 	\sum_{i} P(|i_L i_R\rangle)   	\left[ e^{it eV + i p (x - x')} 	\langle i_L i_R | \psi^\dagger_L(x',t) \psi_R(x',t) \psi^\dagger_R(x,0) \psi_L(x,0)  | i_L i_R \rangle  \right. 
\\&& ~~~~~~~~~~~~~~~ -\left.  e^{-i t e V - i p (x - x')} 	\langle i_L i_R |  \psi^\dagger_R(x',t) \psi_L(x',t) \psi^\dagger_L(x,0) \psi_R(x,0)  | i_L i_R \rangle \right] \\
&=&   -e|g|^2 \int dt \int dx \int dx'  	\left[ e^{it eV + i p (x - x')} 	G_<^L(t,x',x)  G_>^R(t,x',x) -   e^{-i t e V - i p (x - x')}  G_<^R(t,x'x) G_>^L(t,x',x) \right] 
\end{eqnarray*}
where we have defined Green's functions
\begin{eqnarray*} 
	G_<(t,x',x') &=& \langle  \psi^\dagger(x',t) \psi(x,0)\rangle
	\\	G_>(t,x',x) &=& \langle  \psi(x',t) \psi^\dagger(x,0)\rangle
\end{eqnarray*}
where the expectation inludes an expectation over the initial state.  I.e., we really mean
$$
  G_<(t,x',x) = {\rm Tr}[\rho  \, \psi^\dagger(x',t) \psi(x,0) ] 
$$
with $\rho$ the density matrix. 

For the energy current we are going to need the following interesting identity (using the fact that the Green's function only depends on the difference in two times)
\begin{eqnarray*}
\frac{d}{dt}G_<(t,x',x) &=& \frac{d}{dt}\langle  \psi^\dagger(x',t) \psi(x,0)\rangle =  \frac{d}{dt}\langle  \psi^\dagger(x',0) \psi(x,-t)\rangle  
\\ 
&=&  -i \langle  \psi^\dagger(x') [H,\psi(x,-t)]\rangle  = i \langle  \psi^\dagger(x',t) [\psi(x,0),H]\rangle
\end{eqnarray*}
and similarly 
\begin{eqnarray*}
	\frac{d}{dt}G_>(t,x',x) &=& \frac{d}{dt}\langle  \psi(x',t) \psi^\dagger(x,0)\rangle =  \frac{d}{dt}\langle  \psi(x',0) \psi^\dagger(x,-t)\rangle  
\\
	&=&  -i \langle  \psi(x') [H,\psi^\dagger(x,-t)]\rangle  = i \langle  \psi(x',t) [\psi^\dagger(x,0),H]\rangle
\end{eqnarray*}

Using this identity, the similar manipulations give us
\begin{eqnarray*}
J^Q &=&  |g|^2  \int dt \int dx \int dx' 	\sum_{i} P(|i_L i_R\rangle)   	\left[ e^{it eV + i p (x - x')} 	\langle i_L i_R | \psi^\dagger_L(x',t) \psi_R(x',t) \psi^\dagger_R(x,0) [\psi_L(x,0),H_L]  | i_L i_R \rangle  \right. 
\\&& ~~~~~~~~~~~~~~~ +\left.  e^{-i t e V - i p (x - x')} 	\langle i_L i_R |  \psi^\dagger_R(x',t) \psi_L(x,t) [\psi^\dagger_L(x,0),H_L] \psi_R(x,0)  | i_L i_R \rangle \right] \\
&=&   -|g|^2 \int dt \int dx \int dx'  	\left[ e^{it eV + i p (x - x')}  G_>^R(t,x',x) 	\frac{i d}{dt} G_<^L(t,x',x)  +   e^{-i t e V - i p (x - x')}  G_<^R(t,x'x) \frac{i d}{dt} G_>^L(t,x',x) \right] 
\end{eqnarray*}

Let us define the Fourier transform conventions
\begin{eqnarray*} G_<(E,x',x) &=&  \int dt \, e^{-i t E}  \langle  \psi^\dagger(x',t) \psi(x,0)\rangle = \int dt \ e^{-i t E} G_<(t,x',x)
	\\	G_>(E,x',x) &=& \int dt \, e^{i t E}  \langle  \psi(x',t) \psi^\dagger(x,0)\rangle = \int dt \ e^{i t E} G_>(t,x',x)
\end{eqnarray*}
implying the inverses 
\begin{eqnarray*} G_<(t,x',x) &=&   \frac{1}{2\pi}\int dE \ e^{i t E} G_<(E,x',x)
	\\	G_>(t,x',x) &=&  \frac{1}{2\pi} \int dE \, e^{-i t E}   G_>(E,x',x)_.
\end{eqnarray*}
such that 
\begin{eqnarray*} \frac{i d}{dt} G_<(t,x',x) &=&   \frac{1}{2\pi}\int dE (-E) \ e^{i t E} G_<(E,x',x)
	\\	\frac{i d}{dt} G_>(t,x',x) &=&  \frac{1}{2\pi} \int dE \, (E) e^{-i t E}   G_>(E,x',x)_.
\end{eqnarray*}

We then have
\begin{eqnarray} \label{eq:Jeres1}
J^e &=&   \frac{-e |g|^2}{(2 \pi)^2}  \int dt \int dx \int dx'  	\int dE \int dE' \\ \nonumber
& &  \left[ e^{it eV + i p (x - x') + i E t - i E' t} G_<^L(E,x',x)  G_>^R(E',x',x) -   e^{-i t e V - i p (x - x') + i E t - i E' t}  G_<^R(E,x'x) G_>^L(E',x',x) \right]  \\  \nonumber
&=&   \frac{-e |g|^2}{(2 \pi)}   \int dx \int dx'  	\int dE \left[ e^{i p (x - x')} G_<^L(E,x',x)  G_>^R(E+eV,x',x) -   e^{-i p (x - x')}  G_<^R(E,x'x) G_>^L(E-eV,x',x) \right] \\  \nonumber
&=&   \frac{-e |g|^2}{(2 \pi)}   \int dx \int dx'  	\int dE \left[ e^{i p (x - x')} G_<^L(E,x',x)  G_>^R(E+eV,x',x) -   e^{-i p (x - x')}  G_<^R(E+e V,x'x) G_>^L(E,x',x) \right]
\end{eqnarray}
Similarly for the thermal current we obtain
\begin{eqnarray}
	J^E &=&   \label{eq:JEres1}
 \frac{-|g|^2}{(2 \pi)^2}  \int dt \int dx \int dx'  	\int dE \int dE' \\  \nonumber
& &  \left[ e^{it eV + i p (x - x') + i E t - i E' t} (-E) G_<^L(E,x',x)  G_>^R(E',x',x) +   e^{-i t e V - i p (x - x') + i E t - i E' t}  G_<^R(E,x'x) (E) G_>^L(E',x',x) \right] 
\\  \nonumber
&=&	 \frac{|g|^2}{(2 \pi)}   \int dx \int dx'  	\int dE \left[ e^{i p (x - x')}  E G_<^L(E,x',x)  G_>^R(E+eV,x',x) - (E-eV)  e^{-i p (x - x')}  G_<^R(E,x'x) G_>^L(E-eV,x',x) \right] \\ \nonumber
&=&   \frac{|g|^2}{(2 \pi)}   \int dx \int dx'  	\int dE E \left[ e^{i p (x - x')}   G_<^L(E,x',x)  G_>^R(E+eV,x',x) -   e^{-i p (x - x')}  G_<^R(E+eV,x'x) G_>^L(E,x',x) \right]_.	
	\end{eqnarray}

\subsection{Necessary Green's Functions}

\subsubsection{Edge Green's Functions}

We are concerned with the tunneling from an integer edge to the fractional edges in the AntiPfaffian.   We consider the integer edge to be the $R$ system in the above equation and the fractional edges (the Bose plus Majorana edges) to be the $L$ system.   The Green's functions for an integer edge (using our conventions) are simple to calculate, obtaining 
\begin{eqnarray*} 
	G_<(E,x',x) &=&   v^{-1} e^{i (E/v) (x - x')} n_F(E)  
	\\G_>(E,x',x) &=& v^{-1} e^{i (E/v) (x' - x)} n_F(-E)  
\end{eqnarray*}
where $v$ is the edge velocity.  The derivation of these results are not hard and are presented in section \ref{sub:integeredge} below. 

\subsubsection{Factoring the AntiPfaffian Fractional Edge}

\label{sub:factor}

More interesting is to determine the $R$ Green's functions in Eqns. \ref{eq:Jeres1} and \ref{eq:JEres1} above describing the electron when it tunnels into the fractional edge.    

The fractional part of the AntiPfaffian edge contains an upstream Bose $b$ and an upstream Majorana mode $\xi$.  These have the commutation relations
\begin{eqnarray*}
 \{ \xi(x), \xi(x') \} &=& \delta(x -x') \\
\,
 [ b(x), b^\dagger(x') ] &=& \delta(x-x') 
\end{eqnarray*}
The electron operator along this fractional edge is a combination of the Bose and Majorana operators[S2].   
$$
 \psi(x) = \sqrt{\ell_c} \,\, \xi(x) b(x)
$$
Here $\ell_c$ is a cutoff length scale.  For a typical one dimensional system we use a cutoff $2 \pi/q_{max}$ where $\hbar v q_{max} = \Delta$ with $v$ the mode velocity and $\Delta$ the excitation gap energy.  If we think about a system as being on a lattice we should probably instead choose $q_{max}= \pi/\ell$ to match the relationship between the unit cell and the Brillouin zone boundary.   Now in this case, we have an issue that the bose and majorana velocities can be quite different.  As such we will choose the geometric mean
\begin{equation}
\label{eq:cutoff}
 \ell_c = \frac{\pi\sqrt{v_b v_m} }{\Delta}
\end{equation}

The Green's function for the electron operator on the fractional edge can then be written as the product of the Bose and Majorana edges which factorize 
\begin{eqnarray}  \label{eq:Gfactorize}
	G_<(t,x',x) &=&  \langle  \psi^\dagger(x',t) \psi(x,0)\rangle = \ell_c \langle b^\dagger(x',t) b(x,0) \rangle \langle \xi(x',t), \xi(x,0) \rangle = \ell_c G^b_<(t,x',x) G^\xi(t,x',x)  
	\\	G_>(t,x',x) &=& \langle  \psi(x',t) \psi^\dagger(x,0)\rangle = \ell_c \langle b(x',t) b^\dagger(x,0) \rangle \langle \xi(x',t) \xi(x,0) \rangle = \ell_c G^b_>(t,x',x) G^\xi(t,x',x)
\end{eqnarray}
where here we have defined
\begin{eqnarray*} 
	G^b_<(t,x',x) &=& \langle b^\dagger(x',t) b(x,0) \rangle
	\\	G^b_>(t,x',x) &=&  \langle b(x',t) b^\dagger(x,0) \rangle 
\end{eqnarray*}	
and
$$
 G^\xi(t,x',x) = \langle \xi(x',t) \xi(x,0) \rangle
$$
Note, that since for Majorana fields $\xi = \xi^\dagger$ there is no $<$ or $>$ index on this Green's function. 

Defining the usual Fourier transforms
\begin{eqnarray*} 
G^b_<(E,x',x) &=&  \int dt \ e^{-i t E} G^b_<(t,x',x)
	\\	G^b_>(E,x',x) &=&  \int dt \ e^{i t E} G^b_>(t,x',x)
\end{eqnarray*}
and for the Majorana field
\begin{eqnarray}  \nonumber
G^\xi_<(E,x',x) &=&  \int dt \ e^{-i t E} G^\xi(t,x',x) \\
	G^\xi_>(E,x',x) &=&  \int dt \ e^{i t E} G^\xi(t,x',x) = G^\xi_<(-E,x',x) \label{eq:Gmid}
\end{eqnarray}
Using the factorization from Eq.~\ref{eq:Gfactorize} for the electron field Green's function, we obtain the convolutions
\begin{eqnarray} \label{eq:Gconv}
 G_<(E,x',x) &=& \frac{\ell_c}{2 \pi} \int dE' \,\, G_<^b(E-E',x',x)  G_<^\xi(E',x',x) \\ 
 G_>(E,x,x') &=& \frac{\ell_c}{2 \pi} \int dE' \,\, G_>^b(E-E',x',x)  G_>^\xi(E',x',x) \nonumber
 \end{eqnarray}
Note that the meaning of this expression is obviously to divide the energy into the piece going into the Bose mode and the piece going into the Majorana mode in all possible ways.

When we consider the tunneling between edge modes, we will want to use the Green's function for a free boson mode appropriate for the AntiPfaffian edge, but the Majorana Green's function will be calculated in the presence of the tunneling impurity. 

The Bose mode is equivalent to a $\nu=1/2$ bosonic Laughlin edge theory.  The Green's function is the boson-boson correlator, and is given by 
\begin{eqnarray*}
	G_<^b(E,x,x') &=& \frac{E \tilde \ell_c}{2 \pi v_b^2}
	 n_B(E) e^{i (E/v_b) (x - x')} \\  
	G_>^b(E,x,x') &=& \frac{-E \tilde \ell_c}{2 \pi v_b^2} n_B(-E) e^{-i (E/v_b) (x - x') } 
\end{eqnarray*}
with $n_F$ the Fermi function, $v_b$ the Bose mode velocity, and where $\tilde \ell_c = \pi v_b/\Delta$ is a length scale cutoff.   These result are again fairly standard, but are derived in section \ref{sub:boseedge} for completeness. 

Finally we turn to the Majorana Green's function.  In the absence of the impurity we have 
\begin{eqnarray*}
G_<^{\xi,0}(E,x,x') &=& v_m^{-1} 
	 n_F(E) e^{i (E/v) (x - x')} \\
G_>^{\xi,0}(E,x,x') &=& v_m^{-1}  n_F(-E) e^{-i (E/v_m) (x - x') } 
\end{eqnarray*}
where here $v_m$ is the Majorana mode velocity.  These results are also standard, but are derived in section \ref{sub:majoranaedgenoimpurties} for completeness.   Note here we have inserted a superscript 0 to indicate that this is the Green's function in the absence of the impurity.

We next consider plugging these Green's functions into Eq.~\ref{eq:Gconv} to obtain the electron Green's function along the fractional edge, then we use this in Eqs.~\ref{eq:Jeres1} and \ref{eq:JEres1} along with the Green's function of the integer edge.   So long as $p \gg T/v$ and $p \gg eV/v$ , there will no way to have a momentum conserving scattering and both the electrical and thermal conductance between the two edges will be zero, exactly as we expect. 

Now let us consider the effect of the tunneling impurity.   As mentioned in the main text, the effect of the impurity is to incur a phase shift in the Majorana wave as it scatters past the impurity.  The phase shift is given by (a simple scattering problem, see 
section \ref{sec:phaseshift})
$$
  e^{i \varphi(E)} = \frac{E + i E_0}{E -  i E_0}
$$
where $E_0 = \lambda^2/v$ is the coupling energy (called $E_{coupling}$ in the main text).  When evaluating the Green's function $G^\xi(E,x',x)$, this phase shift will have no effect if both $x$ and $x'$ are on the same side of the impurity.   However, if $x$ and $x'$ are on different sides of the impurity, then the Green's function picks up this additional phase.  Thus assuming the impurity is at position $x=0$ we can write 
$$
G_<^\xi(E,x,x')  = G_<^{\xi 0}(E,x,x')  F(E,x,x')
$$
where $G^{\xi 0}_<$ is the unperturbed Green's function and 
$$
 F(E,x,x')= \left\{ \begin{array}{ll}  1  &   x,x' < 0 \\  1 & x,x' > 0 \\ (E + i E_0)/(E -  i E_0)   & x > 0 > x' \\ (E - i E_0)/(E + i E_0) & x < 0 < x'  \end{array}  \right. 
$$
Let us make the assumption that there is no scattering without the impurity as discussed above. (This is not strictly true at nonzero temperature, because even with a very large momentum mimatch there will be some tiny probability that one can have a highly excited state which can scatter, but this is exponentially small so we may ignore this). It is then useful to write
\begin{eqnarray*}
\delta G_<^{\xi}(E,x,x') &=& G_<^\xi(E,x,x')  -  G_<^{\xi 0}(E,x,x') \\
	&=& G_<^{\xi 0}(E,x,x') \,\, \delta F(E,x,x') \\
	&=& v_m^{-1} e^{i (E/v_m) (x - x')} n_F(E)  \delta F(E,x,x')
\end{eqnarray*}
where
$$
\delta F(E,x,x')= \left\{ \begin{array}{ll}  0  &   x,x' < 0 \\  0 & x,x' > 0 \\  2 i E_0/(E -  i E_0)   & x > 0 > x' \\ - 2i E_0/(E + i E_0) & x < 0 < x'  \end{array}  \right._. 
$$
We may correspondingly write the electrons Green's function
$$
\delta G_<(E,x,x') =  G_<(E,x,x') -G^{0}_<(E,x,x')
$$
where again the superscript $0$ indicates no impurity.  Using the factorization of the Green's function in Eq.~\ref{eq:Gconv} we obtain
\begin{eqnarray*}
\delta G_<(E,x,x') &=& \frac{\ell_c}{2 \pi} \int dE'  G^b_<(E - E',x,x') \delta G_<^\xi(E',x,x')
\\
&=&
\,\,\frac{\ell_c \tilde \ell_c}{(2 \pi)^2 v_m v_b^2} \int dE' e^{i [(E-E')/v_b + E'/v_m] (x - x')}  (E-E') n_B^b(E-E') n_F^\xi(E')  \, \delta F(E',x,x')
\end{eqnarray*}
with $\ell_c$ and $\tilde \ell_c$ are the cutoff length scales. Note that we have labeled the Fermi and Bose functions with superscripts $\xi$ and $b$ so that one can see that they correspond to the  two different edges which most generally may not be at the same temperature.   To obtain $\delta G_>$ we can simply use Eq.~\ref{eq:Gmid} obtaining
\begin{eqnarray*}
\delta G_<(E,x,x') &=& \frac{\ell_c}{2 \pi} \int dE'  G^b_>(E - E',x,x') \delta G_>^\xi(E',x,x')
\\
\delta G_>(E,x,x') &=& 
\,\,\frac{\ell_c \tilde \ell_c}{(2 \pi)^2 v_m v_b^2 } \int dE' e^{-i [(E-E')/v_b + E'/v_m] (x' - x)}  (E'-E) n_B^b(E'-E) n_F^\xi(-E')  \, \delta F(-E',x,x')
\end{eqnarray*}

\subsection{Calculating Response}

From Eqs.~\ref{eq:Jeres1} and \ref{eq:JEres1} can write the general expression
\begin{eqnarray*}
	J^\alpha 
&=&   \frac{|g|^2}{(2 \pi)}   \int dx \int dx'  	\int dE X^\alpha \left[ e^{i p (x - x')}   G_<^L(E,x',x)  G_>^R(E+eV,x',x) -   e^{-i p (x - x')}   G_>^L(E,x',x) G_<^R(E+eV,x'x)\right]_.	
\end{eqnarray*}
where $\alpha=e$ or $E$ (for charge current or energy current) and $X^e = -e$ and $X^E = E$

Let us take the $R$-system to be the integer edge and the $L$-system to be the combined fractional edges.    Again assuming that there is no transport in the absence of the impurity we can then write
\begin{eqnarray*}
	J^\alpha 
&=&   \frac{|g|^2}{(2 \pi)}   \int dx dx'  dE \, X^\alpha \left[ e^{i p (x' - x)}   \delta G_<^L(E,x,x')  G_>^R(E+eV,x,x') -   e^{-i p (x' - x)}   \delta G_>^L(E,x,x') G_<^R(E+eV,x,x')\right] \\
&=&   \frac{|g|^2 \ell_c}{(2 \pi)^2}   \int dx dx'  dE dE'  \, X^\alpha \left[ e^{i p (x' - x)}   G_<^b(E-E',x,x') \delta G^\xi_<(E',x,x') G_>^i(E+eV,x,x') -   \right. \\ & & ~~~~~~~~~~~~~~~~~~~~~~~~
\left.  e^{-i p (x' - x)}   G_>^b(E-E',x,x') \delta G^\xi_>(E',x,x')G_<^i(E+eV,x,x') \right] 
\end{eqnarray*}
where $G^i$ means ``integer" edge.  Note that here we can also choose $X=E'$ to determine the thermal current flowing into the Majorana edge mode only.  Plugging in the above results for the Green's functions we obtain
\begin{eqnarray} \label{eq:thiseqJ}
	J^\alpha 
&=&   \frac{|g|^2 \ell_c \tilde \ell_c}{(2 \pi)^3 v v_b^2 v_m}   \int dx dx'  dE dE'  \, X^\alpha(E,E')  (E-E') \\ \nonumber & & \left[ e^{i(-p + (E-E')/v_b + E'/v_m + (E + e V)/v) (x - x')}     n_B^b(E-E')n_F^\xi(E') \delta F(E',x,x') n_F^i(-E - eV)  +   \right. \\ \nonumber & & ~~~~~~
\left. e^{-i(-p + (E-E')/v_b + E'/v_m + (E + e V)/v) (x - x')}    n_B^b(E'-E)n_F^\xi(-E') \delta F(-E',x,x') n_F^i(E + eV)  \right] 
\end{eqnarray}
Note that the exponent of $i (E+ eV)(x-x')$ of the  integer mode has same sign as the $i(E-E')(x-x')$, this is due to the fact that the integer and fractional modes are opposite directed.   Again, $X^\alpha$ can equal $-e, E$ for electrical or thermal current leaving the integer mode or $E'$ for thermal current into the Majorana mode. 

The integrals over $x,x'$ are now simple via
\begin{eqnarray*}
&& 	 \int dx \int dx'  \delta F(E',x,x')  e^{i A (x-x')} =  
\\	&=& \frac{2 i E_0}{E'- i E_0} \int_0^\infty  dx \int_{-\infty}^0 dx'    e^{i A (x -x')}  - \frac{2 i E_0}{E'+ i E_0} \int_0^\infty  dx' \int_{-\infty}^0 dx    e^{i A (x -x')} \\
&=&  \frac{2 i E_0}{E'- i E_0} \frac{-1}{(A  + i 0^+)^2}  - \frac{2 i E_0}{E'+ i E_0}\frac{-1}{(A  - i 0^+)^2}  = \frac{4 E_0^2}{E'^2 + E_0^2}\frac{1}{A^2}
\end{eqnarray*}
where we have ignored the $0^+$ pieces.   This is valid assuming that $A$  is never close to zero.  Indeed, we are assuming that in the exponents in Eq.~\ref{eq:thiseqJ} that the momenum mismatch $p$ is much larger than the $E/v$ for any of the energies and velocities so there is never scattering in the absence of disorder.    As a result we can replace the constant exponent $A$ by $p$  in all occurances, obtaining 
\begin{eqnarray} \label{eq:thiseqJ0}
	J^\alpha 
&=&   \frac{4 |g|^2 \ell_c \tilde \ell_c E_0^2}{(2 \pi)^3 v v_b^2 v_m p^2}   \int dE dE'  \, \frac{X^\alpha(E,E')  (E-E')}{E'^2 + E_0^2} \\ \nonumber & & \left[ n_B^b(E-E')n_F^\xi(E')  n_F^i(-E - eV)  +    n_B^b(E'-E)n_F^\xi(-E')  n_F^i(E + eV)  \right] 
\end{eqnarray}
As we would hope, if all three modes (Bose, Majorana, Integer) are at the same temperature, and if the Voltage is zero, then the expression in brackets in the second line of Eq.~\ref{eq:thiseqJ0} is exactly zero.   Generally, though we should allow for the possibility that there are three different temperatures in the Bose ($b$), Majorana ($\xi$), and integer ($i$) mode. 

Let us assume the voltage and temperature differences between the modes is small, we can then expand the brackets to obtain
\begin{eqnarray} \label{eq:thiseqJ3}
	J^\alpha 
&=&   \frac{2 |g|^2 \ell_c \tilde \ell_c E_0^2}{(2 \pi)^3 v v_b^2 v_m p^2}   \int  dE dE'  \, \frac{X^\alpha(E,E')  (E-E')}{E'^2 + E_0^2}  \left[\frac{  E  (\beta^i - \beta^b) +  E' (\beta^b- \beta^\xi )   + \beta e V}{\sinh(\beta E) - \sinh(\beta E') + \sinh(\beta(E - E')) } \right]
\end{eqnarray}

From the symmetry of the integrand under $E \rightarrow -E$ and $E' \rightarrow -E'$ it is easy to see that we obtain a heat current only for a thermal difference and an electrical current only for a voltage difference.  This agrees with the intuition that there should be no themoelectric effect for dispersionless edges\cite{KaFi96}. 

For both the electric and thermal current from the integer into the Bose mode, we can assume $E_0 \ll T$ in which case 
$$
 \frac{1}{E'^2 + E_0^2} \approx \pi \delta(E') E_0^{-1} 
$$
and we obtain 
\begin{eqnarray}
	J^\alpha 
&=&   \frac{|g|^2 \ell_c \tilde \ell_c E_0}{(2 \pi)^2 v v_b^2 v_m p^2}   \int  dE  \, X^\alpha    \left[\frac{ E( E (\beta^i - \beta^b) + \beta e V)}{2 \sinh(\beta E)} \right]
\end{eqnarray}
with $X^\alpha=-e$ or $E$ for the electrical or energy current.   This yields
\begin{eqnarray} 
	J^e 
&=&   \frac{e^2 \pi^2 | g|^2 \ell_c \tilde \ell_c E_0 T}{4 (2 \pi)^2 v v_b^2 v_m p^2} V    = 
\frac{e^2 | g|^2 \ell_c \tilde \ell_c E_0 T}{16 v v_b^2 v_m p^2} V 
\\
	J^E 
&=&   \frac{\pi^4 | g|^2 \ell_c \tilde \ell_c E_0 T^2}{8 (2 \pi)^2 v v_b^2 v_m p^2} (\Delta T)  =  \frac{\pi^2 | g|^2 \ell_c \tilde \ell_c E_0 T^2}{32 v v_b^2 v_m  p^2} (\Delta T)  
\end{eqnarray}
where we have used $\int dx x/\sinh(x) = \pi^2/2$ and $\int dx x^3/\sinh(x) = \pi^4/4$.  Here $\Delta T$ is the temperature difference between the integer and Bose mode.  Note that in the main text we use the standard definition of conductance in terms of $G_0$ and thermal conductance in terms of $K_0$ which have factors of $h$ rather than $\hbar$. 

The calculation of the thermal current into the Majorana mode is more challenging.  Here we use $X=E'$ in Eqn. \ref{eq:thiseqJ} and we are concerned only with the response to the temperature differnces. Here the limit of $E_0 \rightarrow 0$ is nonsingular.   Taking this limit we have
\begin{eqnarray} \label{eq:thiseqJ4}
	J^{E'}
&=&   \frac{2 |g|^2 \ell_c \tilde \ell_c E_0^2}{(2 \pi)^3 v v_b^2 v_m p^2}   \int  dE dE'  \, \frac{(E-E')}{E'}  \left[\frac{  E [  (\beta^i - \beta^\xi)  - (\beta^b - \beta^\xi)] +  E' (\beta^b- \beta^\xi )}{\sinh(\beta E) - \sinh(\beta E') + \sinh(\beta(E - E')) } \right]
\end{eqnarray} 
the integrals over $E$ and $E'$ can be performed (see section \ref{sec:nastyintegral}
) to obtain
$$
J^{E'} 
=   \frac{|g|^2 \ell_c \tilde \ell_c E_0^2 T}{9  \pi v v_b^2 v_m p^2}  \,
\, [ (T^i - T^\xi) + 2 (T^b - T^\xi) ]_. 
$$

We choose the coupling constant $g$ (an interaction energy scale associated with scattering) to be the gap energy $\Delta$.  As discussed below in sections \ref{sub:boseedge} and \ref{sub:factor} we have chosen
$$
 \ell_c \tilde \ell_c  = \frac{\pi^2 v_b \sqrt{v_m v_b}} {\Delta^2}
$$
so that the coupling constant in the main text is given by
$$
 |\alpha|^2 = |g|^2  \ell_c \tilde \ell_c  = \pi^2 v_b \sqrt{v_m v_b} 
$$

\section{Some further details}

\subsection{Details of  Integrals}

\label{sec:nastyintegral}

There are two integrals we would like to evaluate
$$
I_n=  \int dx dx'  \left( \frac{x }{x'} \right)^n   \frac{(x-x')}{\sinh x - \sinh x' + \sinh (x - x')}
$$
for $n=0,1$. 
We will do the $x$ integral first.   Shift variables $y=x-x'$ and rewrite the $\sinh$ as exponentials.  This allows us to rewrite the required integral as
$$
I_n =  \int dx'  \frac{2}{1 + e^{x'}} \frac{1}{(x')^n }\int dy \frac{y (x' +y)^n }{(e^y + e^{-x'})(1 - e^{-y})}
$$
The integrals over $y$ can be performed using 3.419.2 and 3.419.3 of Ref. [S3] 
$$
\int dy \frac{y^{1+n}}{(e^{-x'} + e^y)(1 - e^{-y})} = \frac{1}{(2 + n)} \frac{[\pi^2 +{x'}^2](-x')^n}{e^{-x'} + 1} 
$$
for $n=0,1$.   We thus obtain
$$
I_n = \frac{1}{1 + 2 n} \int dx'  \frac{1}{1 + e^{x'}}  \left[\frac{\pi^2 + {x'}^2}{e^{-x'} +1} \right]  
$$
Noting that 
$$
\frac{1}{(1+e^{x})(1 + e^{-x})} = -\frac{d}{dx}\frac{1}{1+e^{x}}
$$
the integral  is not difficult giving
$$
 I_n = \frac{1}{1 + 2n} \frac{4 \pi^2}{3} 
$$

\subsection{Some Useful Identities for $G<$ and $G>$}

Here are some general relationships that the Green's functions must obey.  Note that these identities do not rely on translational invariance.  Nor is it required here the the operator $\psi$ is a fermion creation operator. 
$$
G_<(t,x,x') = \langle \psi^\dagger(t,x) \psi(0,x') \rangle = \langle e^{i t H} \psi^\dagger(x) e^{-itH} \psi(x') \rangle
$$
so 
$$
G_<(t,x,x')^* = \langle \psi^\dagger(x') e^{i t H} \psi(x) e^{-itH} \rangle
 = \langle \psi^\dagger(0,x') \psi(t,x) \rangle  = G_<(-t,x',x)$$
Now consider 
$$
G_<(E,x,x') = \int dt e^{-i t E} G_<(t,x,x')
$$
This gives us
\begin{equation}
 G_<(E,x,x')^* = \int dt e^{i t E} G(-t,x'x) = \int dt e^{-i t E} G(t,x'x)  = G_<(E,x',x) \label{eq:identity1}
\end{equation}
Note this equation is true for Fermi and Bose and Majorana correlators.

Assuming thermal equilibrium, we can derive a further relationship between $G_<$ and $G_>$.  \begin{eqnarray*}
 G_<(t,x,x') &=& \langle \psi^\dagger(x,t) \psi(x',0) \rangle \\
 &=& (1/Z) \sum_{nm} \langle n |\psi^\dagger(x,t) |m \rangle \langle m | \psi(x',0) | n \rangle e^{-\beta E_n}  \\
  &=& (1/Z) \sum_{nm} \langle n |\psi^\dagger(x) |m \rangle \langle m | \psi(x') | n \rangle e^{it (E_n - E_m)} e^{-\beta E_n}
\end{eqnarray*}
with $Z=\sum_n e^{-\beta E_n}$ the partition function. 
Fourier transforming we get
\begin{eqnarray*}
 G_<(E,x,x') &=& 
(1/Z) \sum_{nm} \langle n |\psi^\dagger(x) |m \rangle \langle m | \psi(x') | n \rangle  \delta(E - E_n + E_m) e^{-\beta E_n} 
\end{eqnarray*}
Similarly let us calculate
\begin{eqnarray*}
 G_>(t,x',x) &=& \langle \psi(x',t) \psi^\dagger(x,0) \rangle \\
 &=& (1/Z) \sum_{nm} \langle m |\psi(x',t) |n \rangle \langle n | \psi^\dagger(x,0) | m \rangle e^{-\beta E_m}  \\
  &=& (1/Z) \sum_{nm} \langle m |\psi(x') |n \rangle \langle n | \psi^\dagger(x) | m \rangle e^{it (E_m - E_n)} e^{-\beta E_m} 
\end{eqnarray*}
Fourier transforming (note the opposite transform convention)
\begin{eqnarray}
 G_>(E,x',x) &=&  \nonumber
(1/Z) \sum_{nm} \langle n |\psi^\dagger(x) |m \rangle \langle m | \psi(x') | n \rangle  \delta(E - E_n + E_m) e^{-\beta E_m} \\
&=& G_<(E,x,x') e^{\beta E} \label{eq:Gid1}
\end{eqnarray}
Note that this identity put into the expressions Eq. \ref{eq:Jeres1} and \ref{eq:JEres1} show that there is no net electric or thermal current if there is no voltage difference and no temperature difference between the two systems.

\subsection{Integer Edge}
\label{sub:integeredge}

As a warm-up let us calculate the edge Green's function for an integer edge.   We have Dirac fermions with commutations
$$
  \{ \psi(x), \psi^\dagger(x') \} = \delta(x -x')
$$
Assuming a system size of $L$, we have $k$ quantized as $k=2 \pi n/L$.   We then have the Fourier transform
$$
 \psi_k = \frac{1}{\sqrt{L}} \int dx e^{i k x} \psi(x) 
$$
and in reverse
$$
 \psi(x) = \frac{1}{\sqrt{L}} \sum_k e^{-i k x} \psi_k 
$$
The commutations are then
$$
 \{ \psi_k, \psi^\dagger_{k'} \} = \frac{1}{L} \int dx \int dx' e^{i k x - i k' x'} \{ \psi(x), \psi^\dagger(x') \} = \frac{1}{L} \int dx \, e^{i (k -k') x}  = \delta_{k,k'}
$$
We calculate
\begin{eqnarray*} 
	G_<(t,x',x) &=& \langle  \psi^\dagger(x',t) \psi(x,0)\rangle = \frac{1}{L} \sum_{k,k'}  e^{-i k x + i k' x'} \langle \psi^\dagger_{k'}(t) \psi_k(0) \rangle  =  \frac{1}{L} \sum_{k}  e^{i k (x' -  x)} \langle \psi^\dagger_{k}(t) \psi_k(0) \rangle 
	\\	G_>(t,x',x) &=& \langle  \psi(x',t) \psi^\dagger(x,0)\rangle = \frac{1}{L} \sum_{k,k'}  e^{-i k' x' + i k x} \langle \psi_{k'}(t) \psi^\dagger_k(0) \rangle  =  \frac{1}{L} \sum_{k}  e^{i k (x -  x')} \langle \psi_{k}(t) \psi^\dagger_k(0) \rangle 
\end{eqnarray*}
Assuming the system is thermal with zero chemical potential, we then have
\begin{eqnarray*} 
	G_<(t,x',x) &=& \frac{1}{L} \sum_k  e^{i k (x' - x) + i E_k t}  n_F(E_k) 
	\\	G_>(t,x',x) &=& \frac{1}{L} \sum_{k}  e^{i k (x -  x') - i E_k t} n_F(-E_k))
\end{eqnarray*}
with $n_F$ the Fermi function.   We will now specialize to a linear edge dispersion
$
 E_k = v k
$.
Fourier transforming then gives
\begin{eqnarray*} 
	G_<(E,x',x) &=& \int dt \ e^{-i t E} G_<(t,x',x) = \frac{1}{L} \sum_k  e^{i k (x' - x)}  n_F(E) 2 \pi \delta(E - v k)  \\ &=& \int dk \,  e^{i k (x' - x)} n_F(E) \delta(E - vk) = v^{-1} e^{i (E/v) (x' - x)} n_F(E)  
	\\G_>(E,x',x) &=& \int dt \ e^{i t E} G_>(t,x',x)= \frac{1}{L} \sum_{k}  e^{i k (x -  x')} n_F(-E) \delta(E - vk)
	 \\ &=& \int dk \,  e^{i k (x - x')} n_F(-E) \delta(E - vk) = v^{-1} e^{i (E/v) (x - x')}  n_F(-E)  
\end{eqnarray*}
Note that these expressions correctly satisfy Eq.~\ref{eq:Gid1}.   

The integer Green's functions that we have calculated will constitute the $R$ side of the tunneling system in Eqns. \ref{eq:Jeres1} and \ref{eq:JEres1} above.

\subsection{Majorana Edge Without impurity}
\label{sub:majoranaedgenoimpurties}

We start with the Majorana operators
$$
 \{\xi(x) ,\xi(x') \} = \delta(x-x')
$$
Assume the system is of size $L$, and $k$ is quantized as $k = 2 \pi n/L$.   Fourier transforming we get
$$
 \xi_k = \frac{1}{\sqrt{L}} \int dx {e^{i k x}} \xi(x) 
$$
and in reverse
$$
\xi(x) = \frac{1}{\sqrt{L}}\sum_k e^{-i k x} \xi_k
$$
where $L$ is the system size (assumed infinite).  So that
\begin{eqnarray*}
 \{\xi_k, \xi_{k'}\} &=& \frac{1}{L} \int dx \int dx'   e^{i k x + i k' x'}   \{\xi(x) ,\xi(x') \} 
 \\
 &=& 
 \frac{1}{L} \int dx \int dx'   e^{i k x + i k' x'}  \delta(x - x')  \\
&=& 
\frac{1}{L}  \int dx e^{i (k +k') x}  = \delta_{k+k'}
 \end{eqnarray*}
We can thus think of $\xi_k$ with $k > 0$ as Dirac fermion creation operators with the corresponding $\xi_{-k}$ being the annihilation operators.  The vacuum is the absence of any fermions (or  equivalently the negative $k$ states are filled).

In the absence of a localized Majorana, the correlator is
 \begin{eqnarray}
  G^\xi(t,x',x)&=&\langle \xi(x',t) \xi(x,0) \rangle =   \frac{1}{L}\sum_{k,k'} e^{-i k x - i k' x'} \langle \xi_{k'}(t) \xi_{k}(0) \rangle  \label{xx:corr} \\
  &=&  \frac{1}{L}\sum_{k} e^{i k (x' - x)} \langle \xi_{-k}(t) \xi_{k}(0) \rangle  =    \frac{1}{L}\sum_{k} e^{i k (x' - x) - i k v_m t} \langle \xi_{-k} \xi_{k} \rangle  \nonumber
 \\ &=& \frac{1}{L}\sum_{k} e^{i k (x' - x) - i k v_m t} \, (1 - n_F(v_m k))   = \frac{1}{L}\sum_{k} e^{i k (x' - x) - i k v_m t} \,  n_F(-v_m k)  \nonumber \\
 &=& \frac{1}{L}\sum_{k} e^{-i k (x' - x) + i k v_m t} \,  n_F(v_m k)   
 \end{eqnarray}
where $v_m$ is the Majorana mode velocity. Fourier transforming we obtain
\begin{eqnarray*}
 G^\xi_<(E,x',x) &=& \frac{1}{L}\sum_{k} e^{-i k (x' - x)} \, n_F(v_m k)   \int dt e^{-i t E + i k v_m t}  \\ &=& v^{-1}_m n_F(E) e^{-i (E/v_m) (x' - x)}
\end{eqnarray*}
and correspondingly 
\begin{eqnarray*}
 G^\xi_>(E,x',x) &=& G^\xi_<(-E,x',x) = v_m^{-1}n_F(-E) e^{i (E/v_m) (x'-x)}
\end{eqnarray*}
 
\subsection{Bose Edge}
\label{sub:boseedge}

The $\nu=1/2$ Bose mode can be viewed as two seperate Majorana modes[S2].  The boson operator for an edge with velocity $v$ is a product of the two Majoranas having the same velocity $v$.  We thus write
$$
 b(x) = \sqrt{\tilde \ell_c}\,\, \xi_1(x) \xi_2(x)
$$
with $\tilde \ell_c$ a cutoff length scale.    As with the discussion above by Eq.~\ref{eq:cutoff} we should choose this to be 
$$
 \tilde \ell_c = \frac{\pi v_b}{\Delta}
$$
with $v_b$ the bose mode velocity and $\Delta$ the gap.

We can thus have
$$
 G^b_<(t,x',x) = \ell_c \langle \xi_2(x',t) \xi_1(x',t) \xi_1(x,0) \xi_2(x,0) \rangle = \ell_c [G^\xi(t,x',x)]^2
$$
Fourier transforming we have
\begin{eqnarray*}
G^b_<(E,x',x) &=& \frac{\tilde \ell_c}{2 \pi} \int dE' \,  G^\xi_<(E-E',x',x)  G^\xi_<(E',x',x)  \\
&=&  \frac{\tilde l_c}{2 \pi v^2}  e^{-i (E/v)(x-x')} \int dE'  n_F(E - E') n_F(E')
\end{eqnarray*}
The final integral can be performed by elementary methods to give $ E n_B(E)$ with $n_B$ the Bose function.  Thus we have 
$$
G^b_<(E,x',x) = \frac{\tilde l_c}{2 \pi v^2}  e^{i (E/v)(x-x')} \, E \, n_B(E)
$$
and correspondingly we have
$$
G^b_>(E,x',x) =   \frac{\tilde l_c}{2 \pi v^2}  e^{i (E/v)(x'-x)} \,(-E) \, n_B(-E)
$$

\section{Majorana Edge Plus Majorana Impurity Scattering Problem}

\label{sec:phaseshift}
The scattering phase shift problem of a Majorana edge tunnel coupled to a Majorana impurity has been addressed a number of times previously (See Refs.~\onlinecite{Fendley,Roising,Bishara,Rosenow1,Rosenow2} of the main text).  For completeness we give the key steps of the derivation here (in a slightly different language from that of the references). 

We begin with a Hamiltonian density for the Chiral Majorana edge $\xi(x)$ coupled to a trapped Majorana $\gamma$ at position zero
$$
H =  \int dx \,\, \left[ i (v/2) \, \xi(x) \partial_x \xi(x) + i \lambda \xi(x) \gamma \delta(x)  \right]
$$
with $v$ the Majorana velocity and $\lambda$ the coupling strength.  Here $\gamma$  is a Majorana so $\gamma^2=1$ and $\{ \gamma,  \xi(x)\} = 0$.  Note we also have $\{ \xi(x), \xi(x') \rangle = \delta(x - x')$.   

The equations of motion are given by commutations $\partial_t \gamma = i [ H, \gamma]$ and $\partial_t \xi(x) = i [H, \xi(x)]$ which yields
$$
\partial_t \xi(x) = v \partial_x \xi(x) + \lambda \gamma \delta(x) \nonumber
$$
Note that away from $x=0$ this gives the wave equation with velocity $v$.  At $x=0$, keeping singular parts of this equation we get
$$
 \lambda \gamma = -v \left[ \xi(0^+) - \xi(0^-) \right]
$$
And our second equation of motion is
$$
\partial_t \gamma = - \lambda \left[ \xi(0^+) + \xi(0^-)  \right] 
$$
Replacing $\partial_t$ by $-i \omega$ and solving we get
$$
 \xi(0^+)  = \left[ \frac{\omega + i \lambda^2/v}{\omega - i \lambda^2/v} \right]  \xi(0^-)
$$

\subsection{Bound on Fourier Component of Scattering} 
\label{sub:suppedge}

As mentioned in the main text, the tunneling from a Majorana impurity to the edge should be exponential with some decay length $\zeta$.  While no calculations have been made of such couplings, we can use numerical estimates of the decay length of the splitting $E \sim E_{gap} e^{-R/\tilde \zeta}$ between two quasiholes\cite{baraban} separated by a distance $R$ which is $\tilde \zeta = 2.3 \ell_B$.   Since the energy splitting between two putatively degenerate quasihole  states is linear in the matrix element, whereas here we have  $\lambda^2/v$ with $\lambda$ the matrix element we instead obtain $e^{-2 R/\tilde \zeta}$ or a decay length $\zeta  = \tilde \zeta/2 \approx 1.15 \ell_B$.  

The prefactor $\lambda$ has dimensions ${\rm{Energy}} \sqrt{{\rm length}}$ so its natural estimate should then be
$$
 \lambda \approx E_{gap} \sqrt{\ell_B} \, e^{-R/\zeta}
$$
Thus we obtain a coupling energy
$$
 E_{coupling} = \frac{\lambda^2}{v_m} = \frac{E_{gap}^2 \ell_B}{v_m} e^{-R/\zeta} \approx {\rm 1K}\, e^{-R/\zeta}
$$
where we have used $v_m \approx 10^5 {\rm cm}/{\rm sec}$ and $E_{gap} \approx 1$K and $\ell_B = 16$nm.   (We need not be too precise about the prefactor since everything here is dominated by the exponent).  To obtain $E_{coupling} \approx 4$mK, we then have 
$$
 R \approx 5.5 \zeta \approx 6.3 \ell_B
$$
As noted in the main text the smearing of the coupling along the edge should be over a length scale on order $w \approx  \sqrt{R \zeta}$ which is then
$$
 w \approx 3 \ell_B
$$

\section{Edge Equilibration}

\label{sec:edgeequil}

\subsection{Charge Equilibration}

The tunneling current leaving the integer edge at a single impurity is given by 
%
\begin{equation}
\delta j_1 \ = \ - G \, \Delta (\mu_1 - \mu_B) 
\end{equation}
%
Denoting the density of impurities by $n_{\rm imp}$ and considering a piece of the edge with length $\Delta x$, we find for the tunneling current
%
\begin{equation}
\Delta j_1 \ = \ - n_{\rm imp}\, \Delta x \, G\,  (\mu_1 - \mu_B)
\end{equation}
%
Expressing the energy density of  edge mode $i$  as ${1 \over 2 \kappa_i}  \rho_i^2  - \mu_i \rho_i$, we find the relation $\rho_i = \kappa_i \mu_i$. 
Since the current density is given by $j_i = v_i \rho_i$, and since $\kappa_i = {\nu_i \over 2 \pi v_i }$, we find that 
%
\begin{equation}
\mu_i=\pm {2 \pi \over \nu_i}   j_i \ \ .
\end{equation}
%
Here, the $+$-sign applies for the integer mode, and the $-$-sign for the Bose mode. We define $1/\ell_0 = 2 \pi n_{\rm imp} G$.  Note that in the main text we write $G$ in terms of $h$ rather than $\hbar$ absorbing the $2 \pi$.    We then obtain the following differential equation for the the spatial change of the chemical potential of the integer edge mode  
%
\begin{equation}
\partial_x \mu_1 \ = \ - {1 \over \ell_0} \, (\mu_1 - \mu_B) \ .
\end{equation}
%
For the change of the current of the Bose mode, one needs to take into account that  the sign of tunneling current is opposite, 
that the direction of the current is opposite to that of the integer mode, and that the filling fraction is $\nu_B = 1/2$. In total, one finds
%
\begin{equation}
\partial_x \mu_B \ = \ - {2 \over \ell_0} (\mu_1 - \mu_B)
\end{equation}
%
Taking the difference between the differential equations for integer and Bose mode, we finally obtain
%
\begin{equation}
\partial_x (\mu_1 - \mu_B) \ = \ {1 \over \ell_0} (\mu_1 - \mu_B) \ \ .
\end{equation}
%
Introducing the abbreviation $\Delta \mu = \mu_1 - \mu_B$, we can express the solution as
%
\begin{equation}
\Delta \mu(x) \ = \ \Delta \mu(L)\,  e^{(x - L)/\ell_0}  \ .
\end{equation}
%
In addition, the total current $j_1 + j_B$ is conserved, which implies for the chemical potentials
%
\begin{equation}
\mu_1 - {1 \over 2} \mu_B \  \equiv \ \mu_{\rm tot} \ .
\end{equation}
%
We now can express the chemical potentials of integer and Bose edge mode in terms of $\Delta \mu$ and $\mu_{\rm tot}$ as
%
\begin{equation}
\mu_1 \ = 2 \mu_{\rm tot} - \Delta \mu \ , \ \ \ \  \mu_B \ = \ 2(\mu_{\rm tot} - \Delta \mu) \ \ .
\end{equation}
%
We want to impose boundary conditions that the integer mode is injected into the edge at position $x=0$ with chemical potential $\mu_m$, and that the Bose 
mode is injected into the edge with zero chemical potential at spatial position  $x=L$. We then find the solutions
%
\begin{equation}
\mu_1(x) \ = \ \mu_m \left( 1 \ - \ {1 \over 2} e^{(x - L)/\ell_0} \right) \ , \ \ \ \  \mu_B(x) \ = \ \mu_B \left( 1 \ - \ e^{(x - L)/\ell_0} \right) \ .
\end{equation}
giving the equilibration length of $\ell_0$.

\subsection{Thermal Edge Conductance}
\label{sub:thermaledge}

\subsubsection{Two mode model}

Here we assume the heat transferred between the Majorana mode and any other mode is negligible ($E_{coupling}$ very small so $K^{im}$ and $K^{bm}$ are effectively zero), so we can perform the thermal transport calculation by only considering the integer and Bose modes.    Here the energy density per unit length is $k_B (\pi^2/6) T^2/(2 \pi \hbar v)$.   We would write thermal transport equations in terms of the thermal current density as
$$
 \partial_x J_i^Q = n_{imp} K^{ib} (T_i - T_b) 
$$ 
for example where $J_i^Q = (\pi^2/3) T T_i/(2 \pi \hbar)$ where $T_i = T + \mbox{small}$.

We thus have the transport equations
\begin{eqnarray}
\partial_x T_i &=& -\tilde K (T_i - T_b)  \label{eq:thisoneTi} \\
\partial_x T_b &=& -\tilde K (T_i - T_b) 
\end{eqnarray}
where 
$$
 \tilde K = n_{imp} K^{ib}/K_0 
$$
We then have $\partial_x (T_i - T_b) = 0$ so $T_i - T_b$ is a constant and $\partial_x T_i$ and $\partial_x T_b$ are both constants.   We can thus write
\begin{eqnarray}
 T_i &=& T_i^0 - \beta x  \label{eq:Ti0b} \\
 T_b &=& T_b^0 + (L - x) \beta
\end{eqnarray}
where $T_i^0$ is the value of $T_i$ at $x=0$ and $T_b^0$ is the value of $T_b$ at $x=L$, i.e., these are the temperatures in the reservoirs. We thus have 
$$
  T_i - T_b = T_i^0 - T_b^0 - L \beta
$$
Plugging the form of $T_i$ from Eq.~\ref{eq:Ti0b} into Eq.~\ref{eq:thisoneTi} we obtain
$$
 \beta  = \tilde K (T_i^0 - T_b^0 - L \beta ) 
$$
which we solve to get
$$
 \beta = \frac{\tilde K  (T^0_i - T^0_b)}{ 1 + \tilde K L} 
$$
The total heat current (for one edge only) is then 
$$
 J = K_0 (T_i - T_b) =  \frac{K_0 }{1 + \tilde K L}  (T_i^0 - T_b^0)
$$

\subsubsection{Three mode model}

We will now assume that the thermal conductance between the Majorana mode and the integer and Bose mode is small, but not negligible (i.e., $K^{im}$ and $K^{bm}$ are small but not zero).    This is a bit more complicated and we only sketch the solution.   Here we write an equation for all three edges
$$
c^{\alpha}   K_0  \partial_x T^\alpha = 
   \partial_x J^\alpha  = -n_{imp} \sum_\beta K^{\alpha \beta}  T_\beta
$$
where $\alpha,\beta$ are $i$,$b$ or $m$ and $c^\alpha$ is the signed central charge of the three edges $(-1,1,1/2)$ respectively.    Here we take the diagonal components to be 
$$
 K^{\alpha \alpha} = -\sum_{\beta \neq \alpha} K^{\alpha \beta}
$$
so that the full $K$ matrix is taken to be (with rows and columns in the order $i$, $b$, $m$)
$$
 K = \left(
\begin{array}{ccc}
 -\epsilon-1 & 1 & \epsilon \\
 1 & -2 \epsilon-1 & 2 \epsilon \\
 \epsilon & 2 \epsilon & -3 \epsilon \\
\end{array}
\right) K^{ib}
$$
which gives us
$$
M^{\alpha
\beta}= n_{imp}K_0^{-1} (c^\alpha)^{-1} K^{\alpha \beta} = n_{imp} K^{ib} K_0^{-1} \tilde M
$$
with $$\tilde M = 
\left(
\begin{array}{ccc}
 \epsilon+1 & -1 & -\epsilon \\
 1 & -2 \epsilon-1 & 2 \epsilon \\
 2 \epsilon & 4 \epsilon & -6 \epsilon \\
\end{array}
\right)
$$    
which give us the equation
\begin{equation}
\label{eq:maineq}
  \partial_x T^\alpha = 
 -\tilde M^{\alpha \beta} T^{\beta}
\end{equation}
where $x$ is now measured in units of $K_0/(K^{ib} n_{imp})=\ell^b$ the bose relaxation length.   

We thus solve for the eigenvalues $\lambda_j$ and eigenvectors $t_j^\alpha$ of the matrix $n_{imp}K_0 (c^\alpha)^{-1} K^{\alpha \beta}$.    The general solution will be 
$$
 T_\alpha(x)  = \sum_j a_j t_j^\alpha e^{\lambda_j x}
$$
We set the initial conditions of the system to be 
\begin{eqnarray*}
 T_i(0) &=& T_0 \\
 T_b(L) &=& T_1 \\
 T_m(L) &=& T_1 
 \end{eqnarray*}
and solve for the coefficients $a_j$.  The total thermal current (which can be calculated at any position) is $J=\sum_\alpha c^\alpha K_0 T^\alpha$.  While the general expression is rather messy, they can be solved analytically by Mathematica (The precise expression is not enlightening).  


We generally obtain an edge conductance (adding the two additional integer modes, and accounting for heat flowing on both sides of the sample) given by 
$$
 K/K_0 = 2.5  + \frac{2}{1 + A T }   + {\cal O}(\epsilon) 
$$
where $\epsilon = (32/(9 \pi^3)) (E_{coupling}/T)$ is generally small.   We will derive this next.

\subsubsection{Analytic Derivation for small $\epsilon$}

If we assume that $K^{im}$ and $K^{bm}$ are small  we can expand to linear order in these small parameters and obtain analytically simple results.   This is justified by the fact that we have been working to linear order in $E_{coupling}/T$.

We obtain an edge conductance (adding the two additional integer modes, and accounting for heat flowing on both sides of the sample) given by 
$$
 K/K_0 = 2.5  + \frac{2}{1 + A T }   - \epsilon \, C(A T) 
$$
where   $A = L/(l_q^b T)$ and 
$$
 C(x) =  x \,\,\, \frac{2 + 2x + x^2}{ (1 + x)^2} $$
and $\epsilon = (32/(9 \pi^3)) (E_{coupling}/T)$.  Since we will only be concerned with cases where $x=A T > 1$ we can approximate 
$$
 C(x) \approx x
$$
which we use within the main text. 

We now turn to derive this result.   Our approach here will be to first solve the problem in the limit $\epsilon=0$, then treat $\epsilon$ as  a perturbation.   Since we are solving a linear system of equations which is invariant under all $T \rightarrow T + \mbox{const}$  for simplicity we can set $T_0=0$ and $T_1=1$.  Since the equations we need to solve are invariant under shifting all $T$'s by a constant, this will help us avoid carrying around the average temperature. 

In the $\epsilon=0$ limit  (as we have calculated before) we have
\begin{eqnarray*}
    T_i(x) &=& \frac{x}{1 + L} \\
    T_b(x) &=& \frac{1 + x}{1 + L} \\
    T_m &=& 1
\end{eqnarray*}
where here we measure both $x$ and $L$ in units of $\ell^B_q$ is the bose relaxation length (we will continue to do this for simplicity of notation).

We then have our differential equation for $T_m$  (The third line of the matrix Eq.~\ref{eq:maineq})
\begin{equation}
 \partial_x T_m = -2 \epsilon (T_i + 2 T_b - 3 T_M)
 \label{eq:Tmdiff}
\end{equation}
Plugging in the $\epsilon=0$ results for the variables on the right hand side and integrating we get
\begin{equation}
\label{eq:Tmres}
 T_m(x) = 1 + \epsilon \frac{-2 L - 3  L^2 + 2 x + 6  L  x - 3 x^2}{1 +  L}
\end{equation}
Note that this correctly gives $T_m = T_1=1$ at $x=L$.

We still need to find $T_b$ and $T_i$. Let us define 
\begin{eqnarray*}
T_+ &=& T_i + T_b \\
T_- &=& T_i - T_b
\end{eqnarray*}
The first two lines  of the matrix Eq.~\ref{eq:maineq} can be subtracted to give
$$
 \partial_x T_- = \epsilon (T_i + 2 T_b - 3 T_m)
$$
comparing this to Eq.~\ref{eq:Tmdiff} we realize that we have 
$$
 T_- = T_m/2 + C_1
$$
with $C_1$ some constant.   Note that the heat current at $x=0$ is precisely $-K_0 C_1 =  K_0 (T_b(0) + T_m(0)/2)$ since $T_i(x=0)=0$.

The equation for $T_+$ is given by adding the first two lines of the matrix Eq.~\ref{eq:maineq}
\begin{eqnarray*}
 \partial_x T_+ &=& -(2 + \epsilon) T_i + (2 + 2 \epsilon) T_b - \epsilon T_m \\
 &=&  -2 T_-   -  \epsilon (T_i - 2 T_b + T_m) 
\end{eqnarray*}
In the final $\epsilon$ term we can use the unperturbed values of $T_b$ and $T_m$, and for $T_-$ we can use $T_m/2 +C_1$ yielding
$$
 T_+ =  -x  +  \epsilon \frac{2 x + 2 L x + 6 L^2 x - x^2 - 6 L x^2 + 2 x^3}{2(1+L)}  -  2 C_1 x + C_2 
$$
We then have to impose the boundary conditions. First,  we have $T_i(0) = 0$ giving
\begin{eqnarray*}
0  &=&  T_-(0) + T_+(0)  =  T_m(0)/2 + C_1 + C_2  \\ 
0  &=& C_2 + C_1 + \epsilon \frac{-2 L - 3 L^2}{2(1 + L)} + \frac{1}{2}  
\end{eqnarray*}
where in going to the second line we have used Eq. ~\ref{eq:Tmres}. 

Secondly we impose $T_b(0) = 1$, by
$$
 2  = T_+(L) - T_-(L) = -L + \epsilon \frac{L^2 + 2 L + 2 L^3}{2 (1 + L)} - 2 C_1 L + C_2 - (1/2 + C_1)
$$
Subtracting these two equations from each other removes $C_2$ giving
$$
 2 = -2 C_1 (L+1) + \left( -L + \epsilon \frac{4 L + 4 L^2 + 2 L^3 }{2(1 + L)}\right)  - 1
$$
which we can then solve for $C_1$ giving
$$
 C_1 = \frac{3 + L}{2 (1 + L)} + \epsilon \frac{-2 L - 2 L^2 - L^3}{2 (1 + L)^2}
$$
Multiplied by $K_0$ give the thermal current, then we multiply by two to count both sides.   This matches the above quoted result.

\vspace*{10pt}

\hrule

\vspace*{10pt}

[S1] See for example, C.L. Kane and M.P.A. Fisher in Perspectives on Quantum Hall Effects, edited by S. Das Sarma and A. Pinczuk (Wiley, New York, 1997). 

[S2]   M. Levin, B. I. Halperin, and B. Rosenow, Phys. Rev. Lett. 99, 236806 (2007); S.-S. Lee, S. Ryu, C. Nayak, and M. P. A. Fisher, Phys.
Rev. Lett. 99, 236807 (2007).

[S3]  Table of Integral Series and Products, 5ed, I. S. Gradshteyn and I. M. Ryzhik.  Academic Press, (1994).

\end{document}